\newcommand{\be}        {\begin{equation}}
\newcommand{\ee}        {\end{equation}}

\documentclass[article]{mn2e}

\usepackage{epsfig}

\title{Strange stars as persistent sources of gravitational waves}

\author[N. Andersson, D.I. Jones and K.D. Kokkotas]
{N. Andersson$^1$, D. I. Jones$^1$, K.D. Kokkotas$^2$\\
$^1$Faculty of Mathematical Studies, University of Southampton, 
Highfield, Southampton, SO17 1BJ, United Kingdom
\\
$^2$~Department of Physics, Aristotle University of Thessaloniki, Thessaloniki
54006, Greece
}

\begin{document}

\maketitle

\begin{abstract}
We investigate the relevance of the gravitational-wave
driven r-mode instability for strange stars. 
We find that the unstable r-modes affect strange stars 
in a way that is quite distinct from 
the neutron star case. For accreting strange stars
we show that the onset of r-mode instability does not 
lead to the thermo-gravitational runaway that is likely to occur
in  neutron stars. Instead, the strange star evolves
towards a quasi-equilibrium state on a timescale of about a year.
This mechanism could thus explain the clustering of spin-frequencies
inferred from kHz QPO data in Low-mass X-ray binaries. 
For young strange stars we show that the r-mode driven 
spin-evolution is also distinct from the neutron star case. 
In a young strange star the r-mode undergoes
short cycles of instability during the first few months. 
This is followed by a quasi-adiabatic phase where the r-mode
remains at a small, roughly constant, amplitude for 
thousands of years. Another distinguishing feature from the 
neutron star case is that the r-modes in a strange star
never grow to large amplitudes.  
Our results suggest that the r-modes in strange star
emit a persistent gravitational-wave signal
that should be detectable
with large-scale interferometers given an observation time
of a few months. If detected, these
signals would provide unique evidence for the existence
of strange stars, which would put useful 
constraints on the parameters of QCD.
\end{abstract}


\section{Introduction}

Quite plausibly, the most stable form of matter at supranuclear densities 
is a conglomerate of deconfined
strange, up and down quarks. This could have important
consequences for compact stars, with central densities several 
time the nuclear saturation density, that are 
born following supernova explosions \cite{witten}.
It has, in fact, been suggested that if the strange matter hypothesis is
correct then all the observed neutron stars may  be strange stars
\cite{alc86,colpi}. 
If it could be established that strange stars exist in the Universe 
it would have obvious implications for our understanding 
of pulsars. In addition it would put severe constraints on the 
QCD parameters [eg. the bag constant in the MIT model \cite{fjaffe}]. 
It is thus important to establish whether observations can distinguish 
a strange star from a neutron star. Unfortunately, this turns out to be 
a delicate problem. 

Strange stars are held together by both the strong nuclear interaction and
gravity, and the corresponding equation of state is quite accurately
described by uniform density models (see Madsen \shortcite{madrev} for an exhaustive review). 
A consequence of this is that, in
contrast to the neutron star case, very small strange ``dwarfs'' (with mass
$M\sim R^3$) can form.  However, for the canonical mass of $1.4M_\odot$
gravity dominates the strong interaction which leads to strange stars and
neutron stars being similar in size \cite{alc86}.  
In other words, it is not clear that
one would be able to distinguish between the two cases given observed
masses and radii.

One consequence of the extreme stiffness of the strange matter equation of 
state is that strange stars can reach higher rotation frequencies 
than typical neutron stars. 
However, to date the fastest observed pulsar is the 1.56~ms PSR1937+21
which spins at a rate significantly below 
the breakup limit for most proposed supranuclear equation of state. 
In order to indicate the presence of a spinning strange star the 
period would have to be significantly shorter than 1~ms. 
Thus the claimed observation of a 
0.5~ms pulsar in the remnant of SN1987A caused considerable excitement but, 
of course, it turned out to be flawed. Despite several searches
[see for example \cite{edwards}]
no submillisecond pulsars have yet been discovered.

It has been argued that strange stars might be distinguished
by the fact that they cool faster than a neutron star would, but it 
is now accepted that the direct Urca mechanism may be operating [either
by the proton fraction rising above $\sim 10$~\% \cite{lat} or by hyperons 
being present \cite{page}] in the core of a neutron star. 
This would cool the neutron star 
very rapidly, which would make it difficult to identify a
strange star from cooling data.    

Another possible indication of a strange star would be 
thermonuclear flashes from the surface as accreted material 
is converted into strange matter. This mechanism would, however, 
only operate if the strange star was bare. It is generally anticipated
that strange stars will be covered by a thin crust of ``normal'' nuclear
matter \cite{alc86}.  This crust is suspended above the strange matter core
by the electromagnetic field. This means that the crust density cannot exceed
that of neutron drip \cite{gw92}. As neutrons begin to 
drip out of the crust nuclei
they will migrate into the core and be converted
into strange matter. The presence of such a crust
would allow accretion to proceed without necessarily leading to 
flashes from the surface. 

The sole observational indication \underline{against} strange stars is 
provided by the glitching pulsars \cite{alp87}. The standard 
model for explaining 
the large glitches in for example the Vela pulsar relies on 
the transfer of angular momentum between the neutron star crust 
and a superfluid component. It is not easy to see how to construct an
analogous model for glitches in a strange star. 
 
Following the discovery that the r-modes in a rotating perfect fluid star
would be driven unstable via the emission of gravitational waves
[see  \cite{akreview,lindrev} for detailed reviews], Madsen \shortcite{mad98,mad00}
suggested that this mechanism would provide the means for distinguishing
between neutron stars and strange stars.  The gist of Madsen's argument is
that the bulk viscosity of strange matter is several orders of magnitude
stronger than that of normal neutron star matter, which essentially shifts
the window in which the r-modes are unstable to significantly lower
temperatures. As a consequence one would not expect a newly born strange
star to be immediately 
affected by the r-mode instability, while a neutron star might spin
down significantly in the first few months.  An observation of a newly born
pulsar spinning near the Kepler limit might thus provide evidence for the
presence of a strange star.

The realisation that the r-modes in a spinning 
compact star are generically unstable has led to
a considerable effort aimed at understanding the possible
astrophysical relevance of this mechanism.  
In the last two years several crucial issues have been 
investigated. Key results concern the interaction between 
oscillations in the core fluid and the crust \cite{bu,runaway,lou}, 
the role of the magnetic field \cite{spruit,rezz,mendell}, 
superfluidity \cite{lm00,ac01}, and the effect of exotic particles
that are thought to exist in the deep neutron star core \cite{jones,lo01}.  
Of similar importance has been attempts to understand the 
nonlinear saturation of an unstable mode via hydrodynamical simulations
\cite{sf00,ltv01}. 
For a more complete set of references, see recent review
articles \cite{akreview,lindrev} .

In parallell with the various attempts to implement more detailed
physics into the r-mode model, there have been discussions regarding
possible ways that the instability may manifest itself in current
observational data. The original r-mode spin-evolution models
led to results that accord well with the inferred
initial spin period of about 20~ms for the Crab pulsar \cite{lom,aks00}.  
However, the discovery of the 16~ms 
PSR~0537-6910 that is thought to have been born 
spinning significantly faster (at 6-9~ms assuming a reasonable 
braking index) indicates that the naive r-mode model 
needs refinement. Furthermore, one can readily show that external agents 
like the fallback of supernova debris and an acting magnetic
propeller torque are of utmost importance in determining the 
spin of a newly born neutron star \cite{watts}. 

Adopting a scenario first suggested by Papaloizou \& Pringle 
\shortcite{ppinst} and Wagoner \shortcite{wag}, it has been 
suggested that the r-mode instability may be active in mature
accreting neutron stars \cite{akst99}. 
This could provide an explanation for the 
apparent clustering of spin rates (in the range 260-590~Hz)
inferred from kHz QPO data for Low-Mass X-ray Binaries (LMXB).
The original ideas were based on the notion that 
the angular momentum lost through 
gravitational waves radiated by the r-modes would balance the
accretion torque and prevent the star from spinning up.
This would lead to a persistent gravitational-wave 
signal.
However, as was pointed out by Spruit \shortcite{spruit} and Levin
\shortcite{levin}, the onset of instability is likely to
trigger a thermo-gravitational runaway. This leads to a brief 
period (a few months) of r-mode spin-down followed
by a long period (millions of years) of accretion driven spin-up.
Even though the r-mode instability would still be able 
to cause the observed clustering of LMXBs \cite{runaway}, 
the associated gravitational waves would likely not
be detected (as the event rate would be too low).   

As we will show in the following, the r-mode instability
acts rather differently in a strange star.  
We will argue that
unstable r-modes in an accreting strange
star in a LMXB may, in fact, provide a persistent source of gravitational
radiation. We will also consider the evolution of young strange stars.
We will show how the r-mode driven spin-evolution of a strange star
differs significantly from the neutron star case. Most significantly, 
we find that the r-modes never grow to large amplitudes in a strange star. 
The veracity of our results should be testable with large
scale laser interferometers such as LIGO, VIRGO and GEO600 that are about to come
on-line. We thus propose that gravitational-wave astronomy might soon be able
to establish whether the accreting compact stars in LMXBs are, indeed,
strange and perhaps even constrain the parameters of current theoretical
models. 

We need to model the evolution of the spin rate, temperature and 
r-mode amplitude in an accreting strange star. The essential 
elements of our solution to this problem
are discussed in sections~2--3, and we present our results for recycled
strange stars in Section~4. In section~5 we discuss the 
evolution of young strange stars. Our conclusions are in Section~6. 
In an Appendix, we discuss possible refinements to the 
model. In particular, we consider the
deconfinement of the strange star crust as the spin 
of the star changes and how 
nonlinear contributions to a strong viscous dissipation may serve
to saturate an unstable r-mode.

\section{Modelling unstable r-modes in strange stars}

The aim of this section is to collect results from the literature
that will enable us to assess the relevance of the r-mode instability
for, possibly accreting,  strange stars. We have attempted to parameterise 
all relevant relations in order to make the dependence on 
the stellar (and QCD) parameters clear. The various parameterisations
are summarised in Table~\ref{param}.

\begin{table}
\caption{Parameterised variables used in the paper.}
\begin{tabular}{lll}
$M_{1.4}$ & $M/1.4M_\odot$ & mass \\
$R_{10}$ &  $R/10\mbox{ km}$ & radius \\
$P_{-3}$ &  $P/1~\mbox{ms}$ & rotation period \\ 
$T_8$ & $T/10^8~\mbox{K}$ & core temperature \\
$\dot{M}_{-8}$ & $\dot{M}/10^{-8}M_\odot/\mbox{yr}$ & accretion rate \\
$e_{20}$ & $e/20~\mbox{MeV}$ & energy released per accreted nucleon \\
$\alpha_{0.1}$ & $\alpha_s/0.1  $ & strong coupling constant \\
$m_{100}$ & $m_s/100~\mbox{MeV} $ & strange quark mass \\
$\mu_{400}$ & $\mu_d/400~\mbox{ MeV}$ & down quark chemical potential \\
$d_{10} $ & $d/10~\mbox{kpc}$ & distance to gravitational wave source \\
\end{tabular}
\label{param}\end{table}

\subsection{R-mode estimates}

In the rotating frame of reference the r-modes have frequency
\be
\omega_r = { 2m\Omega \over l(l+1)}
\ee
where $\Omega$ is the rotation frequency of the star.

For simplicity we will assume that the star is 
described as an $n=1$ polytrope. The main reason for this is that 
the corresponding r-mode timescales have already been 
summarised by Andersson \& Kokkotas \shortcite{akreview}. 
A more realistic strange star equation of state will be 
stiffer than our model, but we do not expect this to
affect our results significantly.  

Due to the emission of gravitational waves the $l=m=2$ r-mode
(which is  expected to lead to the strongest instability)
grows on a timescale
\be
t_g \approx  - 47 M_{1.4}^{-1} R_{10}^{-4} P_{-3}^{6} \ \mbox{ s}
\label{tgw}\ee
(the minus sign indicates that the mode is unstable).
We  include only the contribution from the
leading current multipole here.

Viscous dissipation tends to suppress the growth of an unstable mode, 
and we find 
\be
t_s \approx 7.4\times 10^7 \alpha_{0.1}^{5/3}
M_{1.4}^{-5/9} R_{10}^{11/3} T_9^{5/3} \mbox{ s}
\label{tsv}\ee
for the shear viscosity. 
For the
bulk viscosity (which is due to the change in 
concentration of down and strange quarks caused by 
the mode oscillation) the 
viscosity coefficient takes the form
\be
\zeta = {\tilde{\alpha} T^2 \over \omega_r^2 + \beta T^4}
\ee
where the coefficients
$\tilde{\alpha}$ and $\beta$ are given by Madsen \shortcite{mad92}. From this 
we see that the bulk viscosity gets weaker 
at both very low and very high temperatures, 
cf. Fig.~1 of Madsen~\shortcite{mad00}. 
For low temperatures  we find that
\be
t_b^{\rm low} \approx 7.9 M_{1.4}^2 R_{10}^{-4}
P_{-3}^2 T_9^{-2} m_{100}^{-4} \mbox{ s} \ .
\label{tbv}\ee
Meanwhile, at high temperatures we cannot readily write down a
parameterised expression for the bulk viscosity timescale.  The relevant
timescale has to be calculated numerically for each given stellar model.
However, this has little impact on our present study since the star 
would likely not remain in the high temperature part of the instability
window long enough for an unstable mode to grow significantly. 
Hence, we do not consider the high-temperature case (above say $10^9$~K)
further in this paper.

\begin{figure}
\vspace*{1cm}
\hbox to \hsize{\hfill \epsfysize=5cm
\epsffile{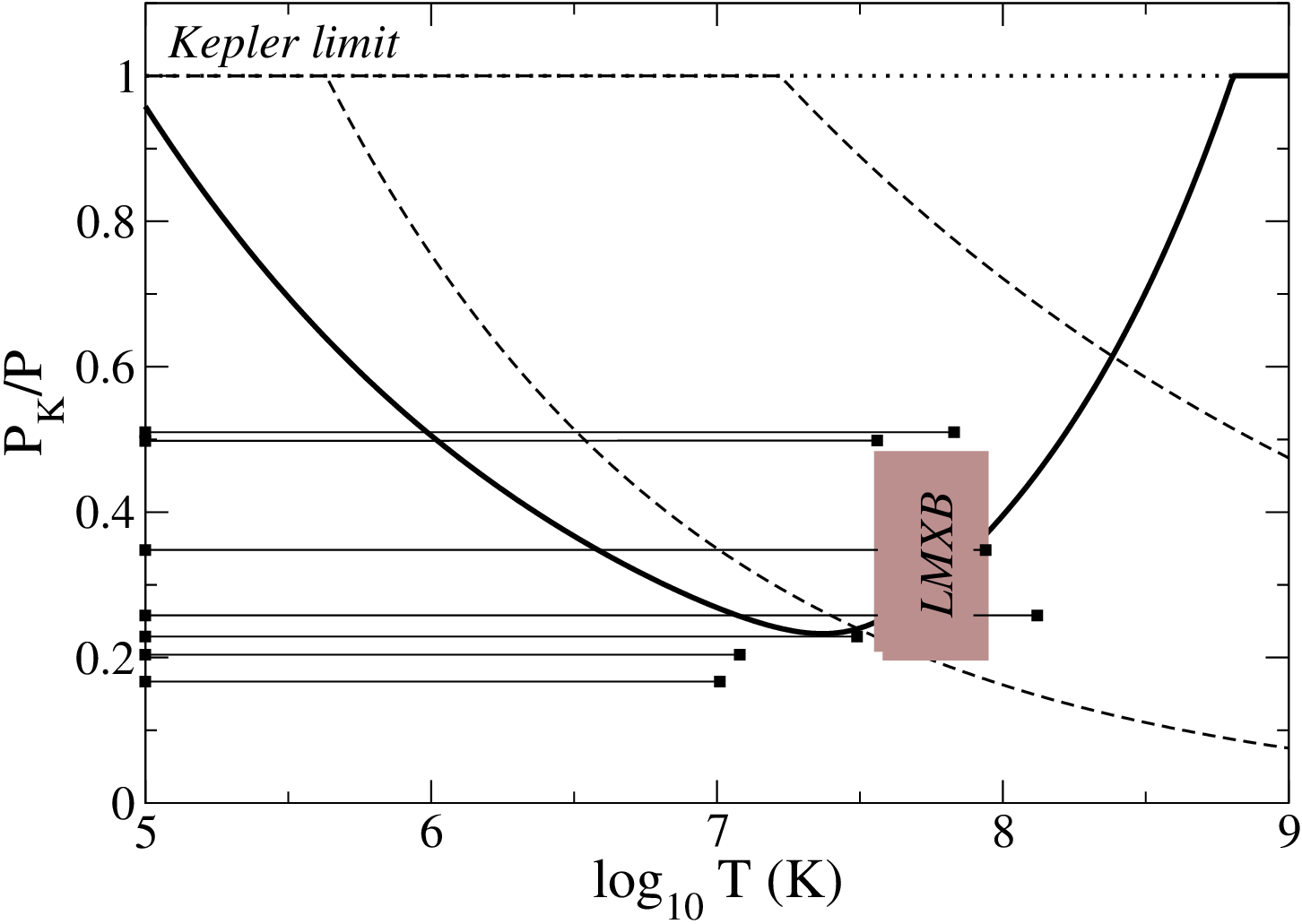} \hfill}
\caption{The r-mode instability window for a strange star
with canonical parameters $M=1.4M_\odot$ and $R=10$~km
(solid line). For comparison we show the corresponding
instability results for normal npe fluid and 
neutron stars with a crust (dashed curves). We compare the 
instability window to i) the inferred spin-periods for accreting stars 
in Low-mass X-ray binaries [shaded box], and ii) 
the fastest known millisecond pulsars 
(for which observational upper limits on the temperature are available) [horisontal lines].
} 
\label{winfig}\end{figure}

Given the above estimates one can readily construct a critical curve
in the $\Omega-T$ plane above which the r-mode will be unstable. 
The mode is unstable whenever
\begin{equation}
\tau^{-1} = t_g^{-1} + t_b^{-1} + t_s^{-1}
\label{instab}\end{equation}
where $t_{g,b,s}$ are the timescales given in Eqns. (\ref{tgw}), (\ref{tsv}) 
and (\ref{tbv}), is negative. If we assume canonical
values for the various parameters
we obtain the results shown in Figure~\ref{winfig} 
[cf. Fig.~1 of Madsen \shortcite{mad00}].

From this figure we immediately see that the r-mode instability would not
become active (apart from a brief initial period expected to last a 
few tens of seconds when the star is significantly hotter than $10^9$~K) until
the star has  cooled to a temperature of a few times $10^8$~K.
A young strange star would therefore not be spun down significantly by the
r-mode instability during its first few months of existence. 
But after a year or so the star will enter the 
``low-temperature'' instability window and undergo a
phase of r-mode spin-down.

\subsection{Temperature evolution}

Given that the viscous damping timescales are strongly dependent
on the temperature we obviously need to model 
the thermal evolution of the star. To do this we first of all
assume that the stellar core is isothermal. 
 
For a strange star, one can estimate that the (integrated) 
heat capacity is \cite{usov01}
\be
C_V \approx 1.5 \times10^{38} M_{1.4}^{2/3} R_{10} T_8 \ \mbox{erg/K}
\ee 
From this we readily infer that the thermal energy of core is
\begin{equation}
\label{eq:Etherm}
E_{\rm thermal} = 7.7 \times 10^{45}  R_{10} M_{1.4}^{2/3} T_8^2 \ \mbox{ erg}.
\end{equation}

At the temperatures that we will consider, the star cools mainly due 
to neutrino emission.  
Pizzochero \shortcite{piz91} gives a formula for the corresponding luminosity
per unit volume, which, when multiplied by the volume of a spherical
star, gives:
\begin{equation}
\label{ednss}
\dot E_{\rm neutrino} = 3.77 \times 10^{37} R_{10}^3 T_8^6 
                                 \rm \, erg \, s^{-1}.
\end{equation}

We can define a cooling rate for an isolated star as
\be
t \approx \frac{2E_{\rm thermal}}{\dot{E}_{\rm neutrino}}
\ee 
(with the factor of 2 included since the thermal energy is 
quadratic in the temperature).
Combining equations (\ref{eq:Etherm}) and (\ref{ednss}) we obtain
\be
t \approx 4.1 \times10^8 M_{1.4}^{2/3} R_{10}^{-2} T_8^{-4} \ \mbox{ s}
\ee
(where we have assumed that the initial temperature is high enough to 
make its contribution to the formula insignificant).
From this formula we see that it would take the star roughly 10 years to 
cool down to a temperature of $10^8$~K. A canonical strange star 
spinning at the Kepler limit enters the main r-mode instability window
in Figure~\ref{winfig} at $T\approx 6\times10^8$~K. Given the above cooling
timescale, this would happen roughly three days after the star was born. 
The star then evolves through the instability phase for the next 
$10^5$ years or so, until the core temperature has fallen below
$10^7$~K. 

If the star is accreting we must also account for heating 
due to the accreted material. In the case of strange stars, the main 
heating source is likely to be due to the conversion of 
 accreted nucleons into strange matter.  
Alcock et al. \shortcite{alc86} estimate a release of $e =
20$ MeV of heat per nucleon due to this conversion.
This then gives
\begin{equation}
\dot{E}_{\rm accretion} = \dot{n} e,
\end{equation}
where $\dot n$ is the rate as which nucleons \emph{enter the strange 
matter core}.
As we will discuss in the Appendix this is 
not necessarily the same as the accretion rate of nucleons onto the star.
In the case where all accreted nucleons are immediately 
converted we get 
\begin{equation}
\label{edass}
\dot{E}_{\rm accretion} = 1.19 \times 10^{37} \dot{M}_{-8} e_{20} 
\, \rm erg \, s^{-1}.
\end{equation}

Finally, a detailed study should also account for the fact that
the shear viscosity will contribute to the heating of the fluid.
We estimate the associated heating rate as
\be
\dot{E}_{\rm viscosity} = 1.1\times 10^{45} \alpha^2 M_{1.4}^{14/9} 
                          R_{10}^{-5/3}  P_{-3}^{-2} \alpha_{0.1}^{-5/3} 
                          T_8^{-5/3}
\ee
where $\alpha$ represents the amplitude of the r-mode (see below).

It is easier to evolve the thermal energy $E_{\rm thermal}$ rather than the
temperature.  
The relevant differential equation is:
\be
\dot{E}_{\rm thermal} = \dot{E}_{\rm accretion} + \dot{E}_{\rm viscosity} - \dot{E}_{\rm neutrino},
\ee

In the absence of an r-mode, the temperature is determined by a balance
between accretion heating and neutrino cooling.  Equations (\ref{edass})
and (\ref{ednss}) then lead to
\begin{equation}
T = 8.25 \times 10^7 \left[ \dot{M}_{-8}  e_{20} R_{10}^{-3}  \right]^{1/6}
         {\, \, \rm K}.
\label{Teq}\end{equation}
For temperatures of this order, shear viscosity dissipation is much smaller
than bulk viscosity dissipation, so that the r-mode instability sets
in when $|t_g| = t_b$.  Equations (\ref{tbv})
and (\ref{tgw}) combine to give the corresponding critical 
rotation period:
\begin{equation}
\label{eq:ptrel}
P \approx 2 T_8^{-1/2} M_{1.4}^{3/4} m_{100}^{-1}
          {\rm \, \, ms}.
\end{equation}

It is illuminating to compare these predictions to the inferred
clustering in the LMXB data. We predict that stars with 
accretion rates in the range $10^{-2} \le \dot{M}_{-8} \le 1$ 
should be confined to spin frequencies in the range $300-450$~Hz. 
Given the many uncertainties in our simple model, this compares
quite favourably with the $260-590$~Hz range suggested by observations.
Conversely, it is clear that
the current observational data cannot be used to rule out 
the possibility that the LMXBs harbour strange stars.

\section{The spin-evolution of accreting strange stars}

In order to investigate the way in which a strange star evolves
under influence of both accretion and an unstable r-mode
we use a phenomenological spin evolution model similar to that
of Owen et al \shortcite{owen}.   We assume
that we can model the star's angular momentum as the sum of the bulk angular
momentum $J$ and the canonical momentum of the r-mode 
$J_c$.  The total
torque on the star $\dot{J}$ can then be written as
\begin{equation}
\label{torque}
\dot{J} = \dot{I}\Omega + I\dot{\Omega} + \dot{J}_c
\end{equation}
where the dots indicate time-derivatives. 
In the following we will write the moment of inertia as $I =
\tilde{I}MR^2$ with  $\tilde{I} = 0.261$ for an
$n=1$ polytrope. 

Note that, since we are modelling the star as an $n=1$ polytrope, accretion
leads to a change in the mass $M$ while the radius $R$ remains constant.
This is a peculiarity associated with our particular stellar model, but it 
has no effect on the outcome of our evolutions. We have verified that 
this is the case by considering other polytropes, and also including the 
effect of the centrifugal flattening in our model. This more general 
study leads to results that are virtually identical to the ones obtained from the 
 model described here so we will discuss only this simple case.   

For the $l=m=2$ r-mode one can show that the canonical 
angular momentum is 
\begin{equation}
J_c = -\frac{3 \Omega \alpha^2 \tilde{J} M R^2}{2}
\label{jcan}
\end{equation}
where $\alpha$ is the mode amplitude (as defined by Owen et al)
and $\tilde{J} = 1.635 \times 10^{-2}$ for an $n=1$ polytrope.

The r-mode is driven by gravitational radiation and damped by viscosity.
We thus assume that the canonical angular momentum evolves according to
\begin{equation}
\frac{\dot{J_c}}{2J_c}= -\frac{1}{\tau}
\label{dotjc}
\end{equation}
(the factor of 2 is included since  $\dot{J_c}$ is proportional to the
square of the perturbation).

The total torque on the star is the sum of torques due to gravitational
radiation and accretion. Throughout the main part of the paper
we  neglect magnetic dipole radiation. There are two simple reasons for this.
Firstly,  
such radiation is likely to be suppressed in an accretion environment,
especially since the magnetic field of the LMXBs are unlikely to be stronger
than $\sim 10^9$ G. Secondly, in the case of young isolated strange stars
one can argue that the main r-mode spindown torque will dominate
the magnetic dipole torque. We will comment on this further in Section~7. 
It is, of course, possible that the interplay between the
r-mode and the interior magnetic field will affect the development of the 
instability (as suggested by, for example Spruit \shortcite{spruit} 
and Rezzolla et al \shortcite{rezz}). We do not account for this 
possibility in our study.  

The
torque from the gravitational wave emission from the $l=m=2$ current
multipole is
\begin{equation}
\label{eq:gwt}
\dot{J}_g = 3 \tilde{J} \Omega \alpha^2 M R^2 t_{g}^{-1} 
\end{equation}
while we assume accretion to lead to a torque
\begin{equation}
\label{eq:acctorq}
\dot{J}_a =  \dot{M} \sqrt{GMR} \ .
\end{equation}
This expression is, of course, likely to be a serious simplification of the
true accretion torque, but it should be sufficient for the rather
qualitative considerations of the present paper.

Equations (\ref{torque})-(\ref{eq:acctorq})  combine to give equations
for the evolution of $\alpha$ and $\Omega$.
\begin{equation}
\label{dotalphaacc}
\dot{\alpha} = - \alpha\left[ \frac{1}{t_g}+
\left( 1 - \frac{3\alpha^2\tilde{J}}{2\tilde{I}}\right)
\left(\frac{1}{t_s} + \frac{1}{t_b}\right)+
\frac{\dot{M}}{2\tilde{I}\Omega}\left({G \over MR^3} \right)^{1/2}\right]
\end{equation}

\begin{equation}
\label{dotomegaacc}
\dot{\Omega} =	\frac{\dot{M}}{\tilde{I}}\left({G \over MR^3} \right)^{1/2}
-{\dot{M} \Omega \over M} - 3\Omega \alpha^2 { \tilde{J} \over \tilde{I}}
\left( { 1 \over t_{\rm sv} } +  { 1 \over t_{\rm bv} }\right)
\end{equation}

The differential equations (\ref{dotalphaacc}) and (\ref{dotomegaacc}) can
be integrated to give trajectories in the $\Omega-T$ plane along which an
accreting strange star would be expected to evolve.

Note that (\ref{dotalphaacc}) indicates that the r-mode is unstable, in the
sense that $\dot{\alpha}>0$, whenever
\be
\frac{1}{\tau}+\frac{\dot{\Omega}}{2\Omega}+\frac{\dot{M}}{2M} < 0
\label{newinst}\ee
This is notably different from (\ref{instab}). 
The difference is, however, easily explained. Eqn
(\ref{newinst}) follows since $J_c$ is conserved for an isolated star, cf.
Ho \& Lai \shortcite{ho}. Eqn (\ref{newinst}) is  the
appropriate criterion for onset of instability in a star that is spun up by
accretion. In our evolutions  the r-mode instability is consequently 
 active whenever (\ref{newinst}) holds.

Equations (\ref{dotalphaacc}) and (\ref{dotomegaacc}) only apply
as long as $\alpha$ remains ``small''. As an unstable r-mode grows
exponentially one would expect the equations to remain valid only for a
very limited period of time. Intuitively, one would expect to 
growth of the mode to be halted once nonlinear effects become relevant. 
Recent attempts to model the evolution of large amplitude 
unstable modes in numerical hydrodynamical simulations suggest
that the r-modes may not saturate until $\alpha$ has grown to values
of order unity \cite{sf00,ltv01}. As we will demonstrate below, the 
r-modes are unlikely to ever grow to such amplitudes in a strange star. 
Thus we assume that the r-mode does not saturate due to nonlinear effects, 
and consider equations (\ref{dotalphaacc}) and (\ref{dotomegaacc})
to be relevant throughout our evolutions.

\section{LMXBs as a source for persistent gravitational waves}

As long as the r-mode remains stable an accreting star will spin up, and
its core temperature should be well approximated by (\ref{Teq}).  As is
clear from Figure~\ref{winfig} this means that the star will enter the r-mode
instability window on the branch where bulk viscosity provides the main
damping mechanism. The subsequent evolution will therefore
be manifestly different
from the neutron star case where the star would undergo a
thermo-gravitational runaway once it reached the instability limit.  We
shall now show that in the case of strange stars the subsequent evolution
may well proceed extremely slowly (on the accretion timescale) and lead to
emission of persistent gravitational waves.
 
As soon as the star enters the instability window the r-mode will grow and
lead to both shear viscosity heating and a spin-down torque.  Intuitively,
one would expect the star to continue spinning up until the r-mode
amplitude becomes sufficiently large to balance the accretion torque.
However, the situation will not be stable unless this balance occurs for a
star residing on the instability curve, i.e. for which we have
$\dot{\alpha}=0$. This situation is well approximated by $|t_g| \approx t_b$.
Use equation \ref{dotalphaacc}) and assume that i) the mode amplitude is
small (such that $1>>3\alpha^2\tilde{J}/2\tilde{I}$) and ii) the
temperature is high enough that the shear viscosity dissipation can be
neglected. 

In order for the situation to correspond to a quasiequilibrium
the mode-amplitude required to balance the accretion torque is equal to:
\be
\alpha^2 \approx 6.7\times10^{-11} \dot{M}_{-8}M_{1.4}^{15/4}R_{10}^{-11/2}
m_{100}^{-7} T_8^{-7/2},
\label{alpa}\ee
while the mode amplitude required for the shear viscosity heating to prevent
the star from cooling down is equal to:
\begin{eqnarray}
\alpha^2 &\approx& 3.6\times10^{-8} M_{1.4}^{-1/18}\alpha_{0.1}^{5/3}
R_{10}^{5/3}m_{100}^{-2} T_8^{2/3} \nonumber \\
&\times& \left[ 3.77R_{10}^3 T_8^6-1.19\dot{M}_{-8}
e_{20} \right] 
\label{alpt}\end{eqnarray}
In obtaining these relations we have made use of (\ref{eq:ptrel})
to express $P$ in terms of $T$.

Now we could in principle equate the two expressions
(\ref{alpa}) and (\ref{alpt}) and solve
for the ``equilibrium'' temperature. However, because this leads to 
a nonlinear equation it is non-trivial. In order to get an 
algebraic answer we instead assume that the star remains close
to the equilibrium temperature (\ref{Teq}). Linearising the 
equation with $T= T_{\rm eq} + \delta T$ it is easy to estimate that 
the required solution corresponds to 
\be
\label{eq:deltaT}
\frac{\delta T}{T_{\rm eq}} = \frac{0.24}
                          { 1 + 942 T_8^{25/6} R_{10}^{43/6} 
                           \alpha_{0.1}^{5/3} m_{100}^5 e_{20} 
                           M_{1.4}^{-137/36}}
\ee
From this we can deduce that once the star enters the instability
regime it never evolves far away from the equilibrium temperature
given by (\ref{Teq}). In other words,  gravitational waves halt the accretion
spin-up once the star reaches the r-mode instability curve.

\begin{figure}
\vspace*{1cm}
\hbox to \hsize{\hfill \epsfysize=5.5cm
\epsffile{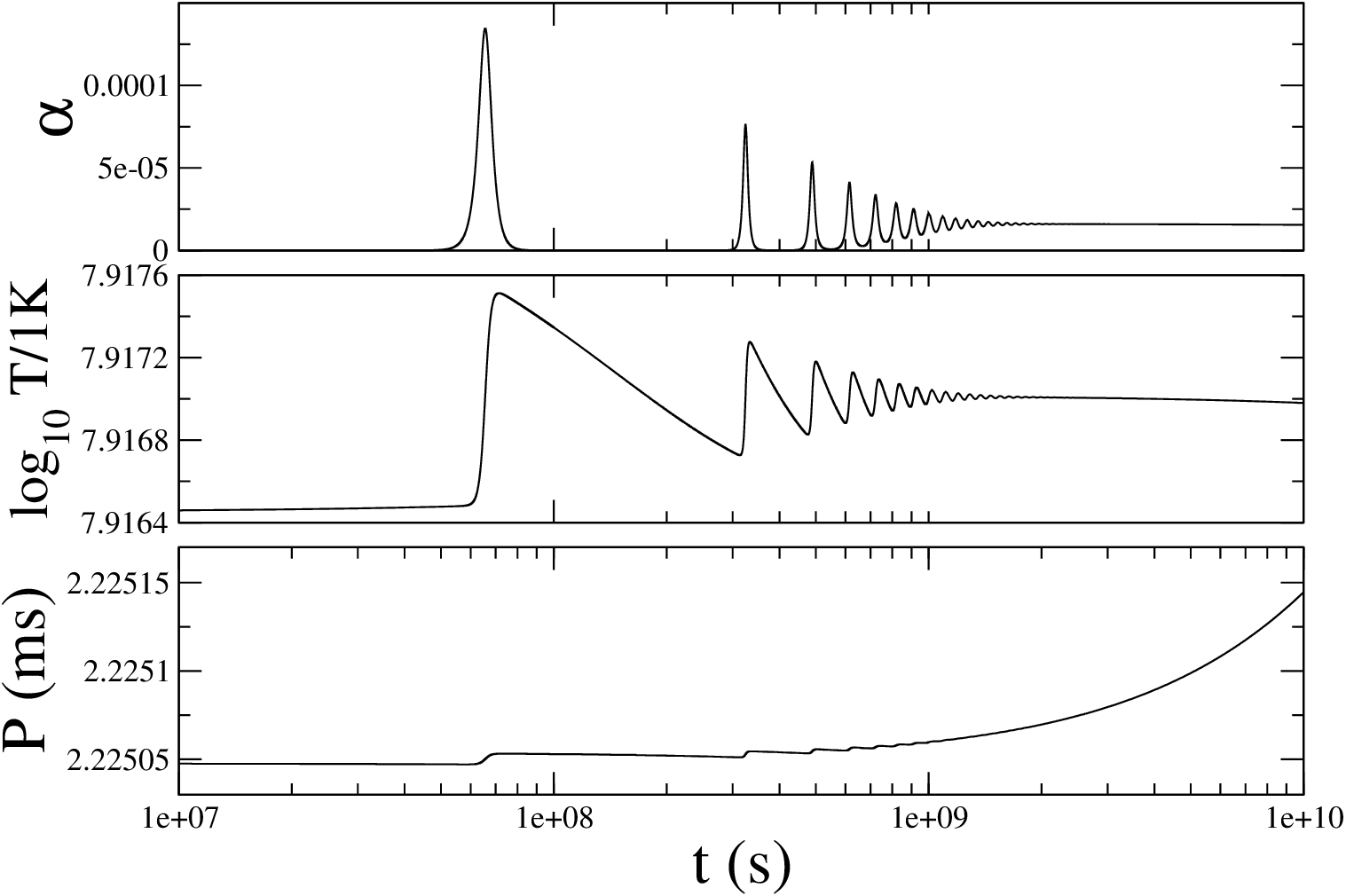} \hfill}
\caption{The evolution of the r-mode amplitude (upper frame), the 
temperature (middle frame) and the spin period (bottom frame) 
corresponding to the first $10^{10}$~s following the initial onset of 
r-mode instability in an accreting strange star.
} 
\label{accrete}\end{figure}

To complete this picture it is necessary to demonstrate that the
equilibrium  described above is a stable one, i.e. that for
small departures from the equilibrium values $(\alpha, T, P)$ the star
returns to the original state.  This is straightforwardly achieved by
assembling our three differential equations for these quantities,
obtaining linearised equations for the departure from equilibrium, and
writing the time dependence of this small departure as $e^{wt}$.  The
solution to the 
eigenvalue problem so obtained shows that the configuration is indeed
stable (the real part of $w$ is negative), with departures from equilibrium being
damped on a timescale intermediate between the temperature evolution
timescale and the gravitational radiation reaction one.  This result will
be of interest should such wandering behaviour ever be seen in an accreting
system. 

Numerical results supporting these analytic estimates are shown in 
Figure~\ref{accrete}. As is readily apparent, after a brief interval 
in which $\alpha$ and $T$ vary sharply
on a timescale close to the thermal one, the system settles
down to a steady state. All subsequent evolution occurs on the 
accretion timescale.

Having established that the spin-up of an accreting strange star 
will be halted by the r-mode instability we can readily assess the 
detectability of the resultant gravitational waves. 
Following Andersson, Kokkotas and Stergioulas \shortcite{akst99} (see 
also Bildsten \shortcite{bild}) we use
\be
h^2 \approx { 4G \over c^3} \left( {1 \over \omega r} \right)^2 |\dot{E}|
\ee
and $\dot{E}=-\omega \dot{J}_{\rm acc} /m$, where $\omega$ is the mode
frequency measured in the inertial frame to give:
\be
h \approx 2.3\times 10^{-27} P_{-3}^{1/2} M_{1.4}^{1/2}R_{10}^{1/4} 
\dot{M}_{-8}^{1/2} d_{10} \ .
\ee
Since the detectability of the source increases roughly as 
the square-root of the number of observed cycles we find that 
the effective amplitude
is 
\be
h_{\rm eff} \approx \sqrt{\omega t_{\rm obs} \over 2\pi} h \sim 10^{-21}
\ee
after  a few months of observation. In other words: this signal should 
easily be detected by a large scale interferometric detector. 

\section{The spin-down of young strange stars}

\begin{figure}
\vspace*{1cm}
\hbox to \hsize{\hfill \epsfysize=5cm
\epsffile{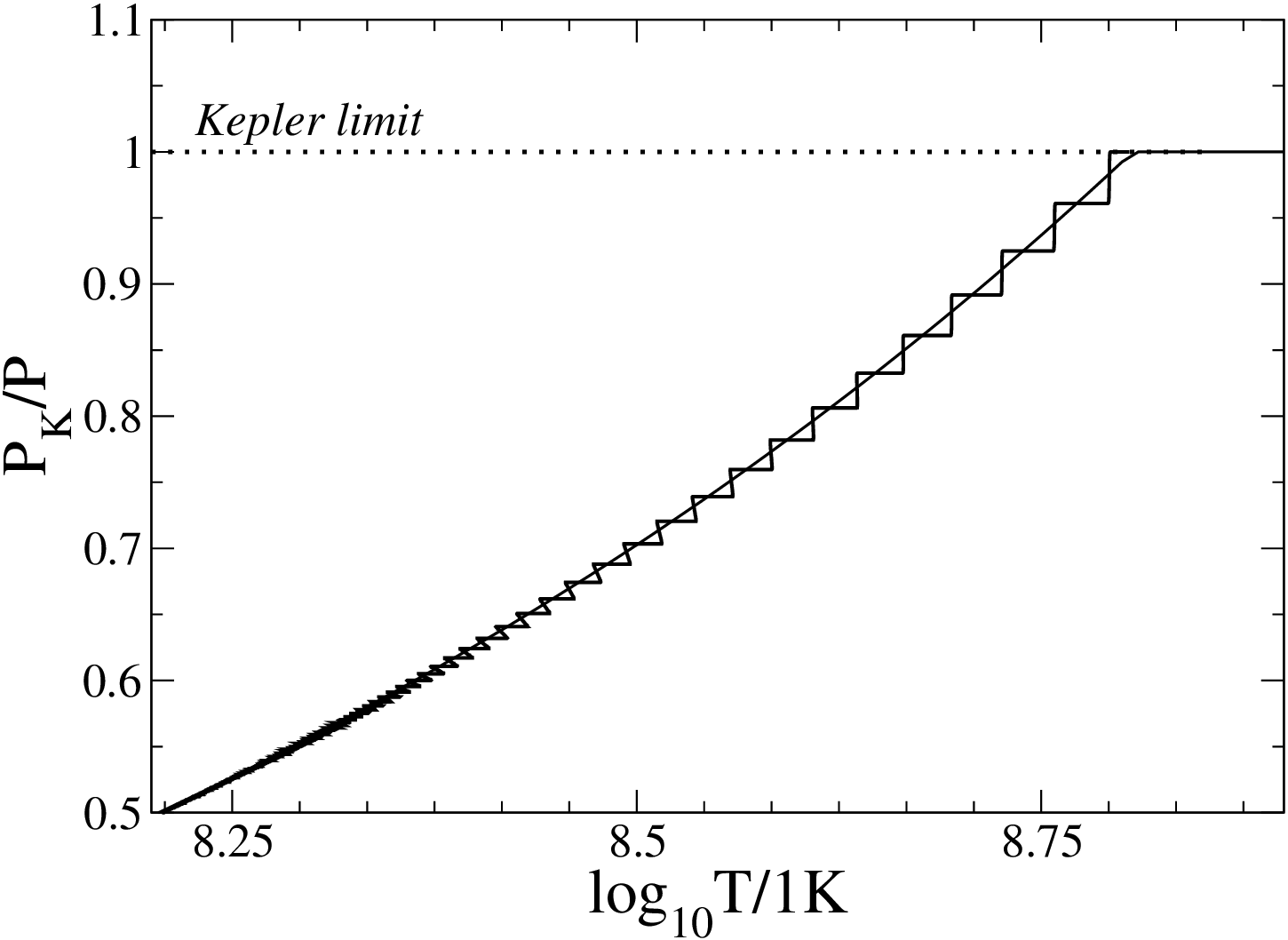} \hfill}
\caption{The evolution of a young strange 
star (with canonical parameters $M=1.4M_\odot$ and $R=10$~km)
under the influence of the r-mode instability.
The thin solid line represents the relevant part of the 
instability window from Fig.~\ref{winfig}. The first year
of the stars evolution following the first onset of instability
are shown as a thick solid line. 
} 
\label{youngevol}\end{figure}

We next want to investigate whether the 
spin-evolution of a young strange star is also distinct from that 
of a nascent neutron star. 
To do this we simply assume that $\dot{M}=0$ in the various 
evolution equations from Section~3 and initialize the integration with 
$P=P_K$, i.e. the star is spinning at the break-up limit
when it cools to  the  temperature at which the r-mode 
first goes unstable. A typical result of this calculation, 
corresponding to the 
first year of the evolution, is 
shown in Figure~\ref{youngevol}. 

At first sight, the result shown in Figure~\ref{youngevol}
may be a bit surprising. A strange star clearly  
evolves in a way that is rather different from the familiar 
results for hot young neutron stars, cf.
Owen et al \shortcite{owen}.  In particular, the strange star never 
evolves far into the instability window for the $l=m=2$ r-mode. 
This is essentially because  the main r-mode 
instability phase does not begin until a strange star has cooled to
a temperature below $10^9$~K. At that temperature the cooling time 
is considerably longer than the spindown time due to an r-mode
of sizeable amplitude. This means that, once the r-mode begins to grow  
it efficiently spins the star down through the 
bulk viscosity part of the instability curve. Once stable, the mode
dies away exponentially and cooling will again govern the 
evolution of the star. Eventually the mode is again unstable and the 
cycle is repeated. After a few years the system reaches a quasi-equilibrium 
where the mode amplitude evolves on the cooling time scale, and the star 
spins down along the bulk-viscosity branch on the instability window. 
That this would be the case has previously been suggested by Madsen 
\shortcite{mad00}.

\begin{figure}
\vspace*{1cm}
\hbox to \hsize{\hfill \epsfysize=5.5cm
\epsffile{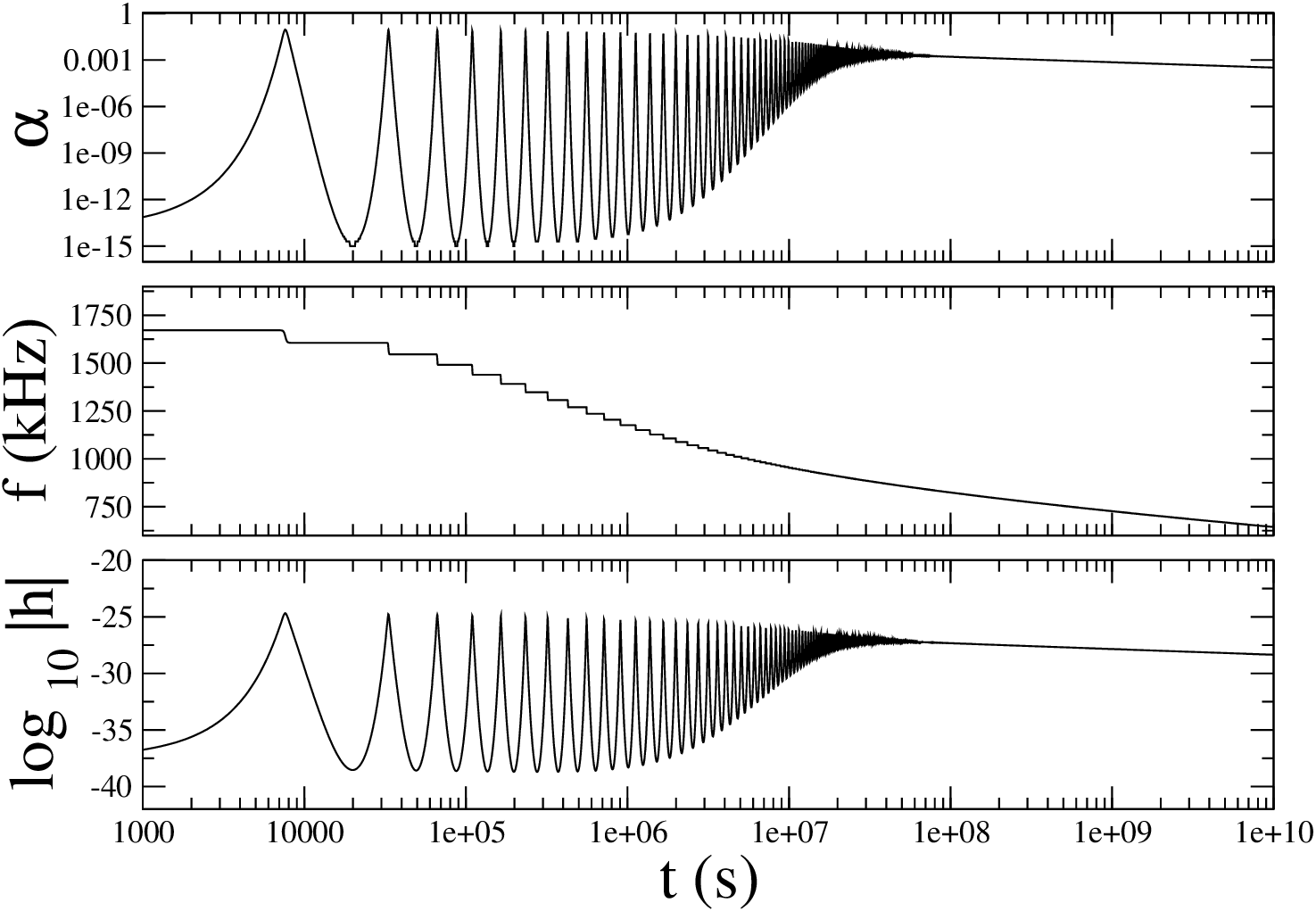} \hfill}
\caption{The upper frame shows the
evolution of the r-mode amplitude during the first year of 
evolution of a young strange star. It can be seen that 
the mode never reaches amplitudes of order unity, and also that 
 the oscillating evolution in the amplitude is replaced by 
a smoother behaviour with small values ($\sim 10^{-2}$) of $\alpha$
after a year or so. The middle frame shows the corresponding evolution of the gravitational-wave
frequency, while the bottom frame illustrates
the gravitational-wave amplitude for a source in the Virgo cluster 
(at a distance of 15~Mpc).
} 
\label{amplitude}\end{figure}

This result has considerable repercussions for the r-modes
as a gravitational-wave source. In Figure~\ref{amplitude}
we show the evolution of the r-mode amplitude corresponding 
to the  case shown in figure~\ref{youngevol}. Here it is 
interesting to note that the mode amplitude varies greatly 
during the initial phase of r-mode spindown. Furthermore, it is clear that
the evolution never leads to $\alpha$ reaching values of order
unity. This is intriguing given the results of 
recent hydrodynamical simulations that seem to suggest that an unstable 
r-mode will not saturate due to nonlinear effects until $\alpha \sim 3$. 
In other words, the r-modes in a strange star may never actually
reach the truly nonlinear regime. If this is, indeed, the case our
spindown model may be better than we initially had any reason to expect.


We now consider the implications of the above results 
for the r-modes as a source of detectable gravitational waves. 
We estimate the gravitational-wave strain using
[cf. Andersson \& 
Kokkotas \shortcite{akreview}],
\be
h \approx 7.54 \times10^{-23} \alpha \tilde{J} P_{-3}^{-3} M_{1.4} 
R_{10}^3 { 15 \mbox{ Mpc} \over D}
\ee
where $D$ is the distance to the source (here taken to be in the 
Virgo cluster). 
Typical results are shown in Figure~\ref{amplitude}.
The  oscillatory evolution of the r-mode amplitude
would, obviously, imprint a unique signature on the 
emerging gravitational waves. In contrast to the neutron star 
case, where the r-mode signal evolves slowly on the cooling timescale,
the signal from a strange star would be concentrated in 
relatively brief ``bursts''. Even though one must recognize the
difficulties in providing a detailed model for this behaviour 
(given the 
need for a detailed, probably nonlinear, understanding of 
for example the viscosities) it seems clear that observations
ought to make it quite easy to distinguish a strange star 
from a neutron star.   

We will not attempt to assess the detectability of the early 
phase of the gravitational-wave signal shown in Figure~\ref{amplitude}. 
The rapid (model dependent!) variations in the mode amplitude 
implies that this problem  requires a more detailed signal 
processing discussion. However, this  initial phase 
last only for a few years. Once the star reaches the 
quasi-adiabatic phase where $\alpha$ remains constant over
long periods in time we can easily estimate the detectability
of the gravitational waves. To do this we consider
 the data in Figure~\ref{amplitude}. 
From the figure  
which we infer that
$h \sim 10^{-27}$  and $f \sim 900~\mbox{Hz}$  at $t \sim 10^8~\mbox{s}$ .
Then, knowing that the effective amplitude improves roughly as the squareroot 
of the number of detected cycles, i.e. that $h_{\rm eff} \approx \sqrt{f t_{\rm obs}} h$, we find that one would need an integration time of 
$t_{\rm obs}\sim40$~years 
to achieve an effective amplitude of order $10^{-21}$. This is 
clearly not feasible, and we can conclude that the late part of the signal
shown in Figure~\ref{amplitude} is unlikely to be detected from sources in the 
Virgo cluster.    

The usual rationale for considering sources in the Virgo cluster is that 
we need a reasonable event rate to make a strong case that a 
certain model leads to detectable gravitational waves.
In the present case, it may be sufficient to focus on sources in 
our own Galaxy. After all,  the gravitational-wave signal from a strange star
lasts considerably longer than that from a young neutron star. 
From Figure~\ref{amplitude} we can see that the signal remains
above $h\sim 10^{-29}$ for more than 300 years. Taking the corresponding 
wave frequency to be 600~Hz (again extrapolated from the figure), we find that 
the effective amplitude would reach $10^{-21}$ after two weeks of integration
if the source was located at a distance of $10$~kpc, i.e. at the centre of 
the Galaxy. Finally combining the expected supernova rate on 3/yr/galaxy
with a lifetime of the gravitational wave signal of 300 years or so, we
deduce that there could be as many as 10 or so systems active in the Galaxy at any given 
time. At least if all compact stars born in supernovae are strange.

The above results offer  exciting possibilities
given the fact that new large scale interferometers
are about to come into operation.  Our study suggests 
that these detectors should be able to test 
the idea that ``all neutron stars are strange stars''
in the near future.

Before we conclude our discussion of young strange stars, it is 
worthwhile making two remarks. Firstly, 
one might wonder whether the r-mode driven spin-down would have 
observational effects
in addition to the generated gravitational waves.
It could, for example, leave a distinct imprint on the electromagnetic 
pulsar signal. After all, the r-mode 
phase may last for thousands of years in a young strange star
born spinning at, or near, the Kepler limit and it is possible that 
the presence of the gravitational-wave torque could lead to some 
anomalous spin-down behaviour. Unfortunately, 
one can argue that this is unlikely to be the case, unless 
the star is observed in the first 10 years or so. 
One can easily show that the r-mode spin-down torque
is dominated by the standard contribution
from an electromagnetic dipole, 
\be
\dot{J}_{\rm em} = - { 2B^2R^6\Omega^3 \over 3c^3} \ ,
\ee 
whenever
\be
B > 1.7 \times 10^{15} \alpha M_{1.4}P_{-3}^{-2} G
\ee
Taking the spin-period to be 2~ms and assuming a magnetic field 
of $10^{12}$~G (fairly typical for a young pulsar) we find that the 
spin-down will be mainly electromagnetic once the mode-amplitude 
falls below $\alpha \sim 10^{-3}$. From Figure~\ref{amplitude} we
can see that this will happen after the first ten years or so.
In other words, we should not necessarily
expect the presence of a small amplitude r-mode 
to make an imprint on the electromagnetic signal from a 
young pulsar.

Secondly, we note a recent paper by Middleditch et al \shortcite{middle}
which  provides possible evidence for the existence of a 2.14~ms pulsar 
in the remnant of SN1987A. The observations were made over several 
years between 1992 and 1996, and the data indicate a
surprisingly large, and 
highly variable spin-down rate. It is interesting to note that
these suggestions agree qualitatively with the results of our simple spin-evolution
model for a newly born strange star. In particular, it is clear that 
one would expect to find a highly variable $\dot{P}$ during the 
phase where the r-mode amplitude oscillates on the cooling timescale.
Should the Middleditch observations prove to be reliable,
they may thus indicate that a strange star was born in SN1987A.

\section{Summary}

In this paper we have studied the r-mode instability
in the context of strange stars. We have shown that 
 unstable r-modes affect strange stars 
in a way that is quite distinct from 
the neutron star case. For accreting strange stars, 
the onset of r-mode instability does not 
lead to a thermo-gravitational runaway. Instead, the strange star evolves
towards a quasi-equilibrium  on a timescale of about a year.
 This  could
explain the clustering of spin-frequencies
inferred from kHz QPO data in Low-mass X-ray binaries. 
For young strange stars we showed that the r-mode driven 
spin-evolution is also distinct from the neutron star case. 
In a young strange star the r-mode undergoes
short cycles of instability during the first few months. 
This is followed by a quasi-adiabatic evolution where the r-mode
remains at a small, roughly constant, amplitude for 
perhaps as long as $10^5$ years. Another interesting 
feature of these evolutions is that the r-modes in a strange star
never grow to large amplitudes. This could prove to be a 
crucially important observation since recent hydrodynamical simulations 
\cite{sf00,ltv01} indicate that the r-modes will not be significantly
affected by nonlinear effects until at much larger amplitudes 
than those reached in our study.   

The main conclusion of this study is that
 the r-modes in a strange star
should emit a persistent gravitational-wave signal
that ought to be detectable
with large-scale interferometers given an observation time
of weeks to months. If detected, these
signals would provide unique evidence for the existence
of strange stars in the Universe, which would put useful 
constraints on the parameters of QCD.

\section*{Acknowledgements}

We  thank John Miller and Nick Stergioulas
for enlightening discussions. NA is a Philip Leverhulme prize
fellow. NA and DIJ acknowledge  
support from PPARC via grant number PPA/G/S/1998/00606.
This work was also supported by the  EU programme 
``Improving the Human Research Potential and the Socio-Economic
Knowledge Base'' (research training network contract HPRN-CT-2000-00137).

\section*{Appendix: Two complicating issues}

In this Appendix we discuss briefly two issues that 
could prove to be relevant for the spin-evolution 
of a strange star. A preliminary study shows that these
effects will not alter the conclusions drawn in this paper
qualitatively, but that they must be considered in 
any attempt to obtain reliable quantitative estimates. 

\subsection{Deconfinement of the crust}

As mentioned in the Introduction, a strange star is likely to 
have a thin crust of normal nuclear matter. The inner edge of this 
crust is sharply defined by the nuclear drip density, since 
the crust relies on the electromagnetic field to suspend it 
above the strange matter core and save it from deconfinement. 
Neutrons are obviously not affected by the magnetic field and
will fall into the strange core as soon as they begin to drip 
out of the nuclei. For a given equation of state, one can 
 determine the maximum mass of the crust.   
This has been done by Glendenning and
Weber \shortcite{gw92}. Not very surprisingly, the maximum mass depends
also on the rotation rate of the star.  
Yuan and Zhang  \shortcite{yz99} have provided a simple analytic
calculation of the maximum crust mass, which agrees reasonably
well with the numerical results of Glendenning and Weber. For simplicity, we
will use the analytic result here, i.e. we assume that
\begin{equation}
M_{\rm crust, max} \approx [ 2.2 + 2.9 R_{10}^{-5}M_{1.4}^{-1}  
              (\Omega/\Omega_{\rm K})^2] \times 10^{-5} M_\odot.
\label{mcrust}\end{equation}
Clearly, this estimate may have impact on a discussion of
accreting strange stars. 

Strange star accretion can be seen as a 
two-stage process. First, the accreted matter, of mass say $dM$, 
is added to the crust.  If $M_{\rm
crust} < M_{\rm crust, max}$ the increment $dM$ can be added
to the crust without any matter migrating into the core
and thus no significant
accretion heating. This is particularly relevant since
accretion  leads to  spin-up and an increase in $M_{\rm crust, max}$

Conversely, in the case of a rapidly spinning
down star (perhaps due to an unstable r-mode), 
a substantial portion of the crust
will `fall' into the core on the spindown timescale.  This may then rapidly
heat the core. Consider the case of spindown due to a saturated
r-mode, for which 
\begin{equation}
\dot{\Omega} \approx { 3 \alpha^2 \Omega \tilde{J} \over \tilde{I} t_g} \ .
\label{dotomegasat}\end{equation}
Taking the time-derivative of (\ref{mcrust}) we have
\be
\dot{M}_c \approx 5.85\times 10^{-5} R_{10}^{-8}M_{1.4}^{-1} { \Omega 
\dot{\Omega} \over \Omega_K^2} \, M_\odot/\mbox{s}
\ee
or, after using (\ref{dotomegasat}), 
\be
\dot{M_c} = -1.5\times10^{-7} \alpha^2 R_{10}^{-1} M_{1.4}^{-1} P_{-3}^{-8}
 \, M_\odot/\mbox{s}
\ee
For large mode amplitudes and a spin-rate near the Kepler limit, this 
would correspond to a huge rate of deconfinement of the crust. 
The effect of this on the temperature of the star, and the evolution 
of the unstable r-mode could clearly  be significant.

The above discussion does, however, come with a few caveats.
The electromagnetically supported gap between the crust and core is
\emph{incredibly} thin: Alcock et al.\ (1986) suggest $10^{-13}$~m, which is
of order $10^{-3}$ atomic diameters, or 10 nuclear diameters. 
This means that,  unless the
crust and core execute nearly exactly the same oscillation the two will
overlap with part of the crust ``dissolving'' into the core.  In the limit
where the crust is infinitely rigid and so does not partake in the
oscillation, we can imagine the interior oscillating core dissolving the
inner part of the crust, in a shell of thickness corresponding to 
the radial motion of
the strange core.  The crust would then presumably rapidly contract, with
another shell being dissolved, and so on.
Of course, in reality
the crust is far from rigid.  Being much thinner and less
massive than a neutron star 
crust, its modulus of rigidity is very small.  It follows
that the crust may participate in a fluid oscillation
mode to a considerable extent,
[as  discussed in the neutron star 
case by Levin \& Ushomirsky (2000)]. 
We have not attempted to quantify the extent to which the 
crust of a strange star participates in (say) a
core r-mode, but it is an interesting question that may be 
worth further consideration. 

Finally, the fact that the core-crust boundary  
corresponds to the density at which neutrons being to drip 
out of the nuclei implies that possible viscous Ekman 
layers will be much less significant than in the neutron star 
case, cf. the comments by Madsen~\shortcite{mad00}.
In view of this, we have not included effects due to viscous
rubbing at the crust interface in our analysis.

\subsection{Nonlinear viscosity saturation}

The possibility that nonlinear contributions to the viscosity
coefficients may saturate an unstable mode has been 
suggested by Reisenegger (private communication).
However, in the case of a ``normal''
neutron star fluid  the standard bulk viscosity 
is too weak for this effect to be relevant (the mode does saturate, but 
only at unphysical amplitudes). In the 
case of strange stars, the bulk viscosity is significantly
stronger and thus it could well be that the nonlinear 
contributions to the viscosity turn out to be relevant.  
Incidentally, it may be worth pointing out that a similar 
effect may be highly relevant in the case of strong hyperon induced
bulk viscosity \cite{jones,lo01}.

We want to investigate the consequences of the fact that the 
low temperature
bulk viscosity coefficient can be written (cf. eqn.~(21) of Madsen
\shortcite{mad92})
\be
\zeta = {\tilde{\alpha} T^2 \over \omega_r^2} 
\left[ 1 + \frac{3}{16\pi^2} \left( {m_s^2 \Delta v \over 3  \mu_d v_0 T
}\right)^2 \right] 
\label{fullbv}\ee
In the estimate used earlier in the paper (eg. to create the data 
displayed in Figure~\ref{winfig}) the $\Delta v /v_0$ term was neglected.
 The bulk viscosity  calculation could, of course, be generalised in 
order to include this  compression term but before doing this it is useful 
to estimate whether this contribution will ever be relevant.
To do this we use
\be
{ \Delta v \over v_0} \approx { \delta r \over R}
\ee  
where $\delta r$ is the radial component of the r-mode
displacement vector. 
The two terms in the bracket of (\ref{fullbv}) 
contribute equally when 
\be{\delta r \over R} \approx 0.07 \alpha \left( 
{ \Omega \over \Omega_K}\right)^2 \approx
7.5\times10^{-2} \mu_{400} m_{100}^{-2} T_9 
\ee
where we have used the  estimate of the height of
the r-mode at the surface of the star provided by 
Andersson and Kokkotas \shortcite{akreview}. 
Parameterising to typical  values 
we can use this result to get
an estimate of the r-mode amplitude at which the nonlinear
contribution to the bulk viscosity should  not
be neglected. We then get
\be
\alpha \approx 1.07 T_9 \mu_{400} m_{100}^{-2} \left({P \over P_K}\right)^2 
\ee
from which we can deduce that the contribution may be important in stars 
cooler than (say) $10^9$~K, i.e. in the main instability of Figure~\ref{winfig}.

\begin{figure}
\vspace*{1cm}
\hbox to \hsize{\hfill \epsfysize=5cm
\epsffile{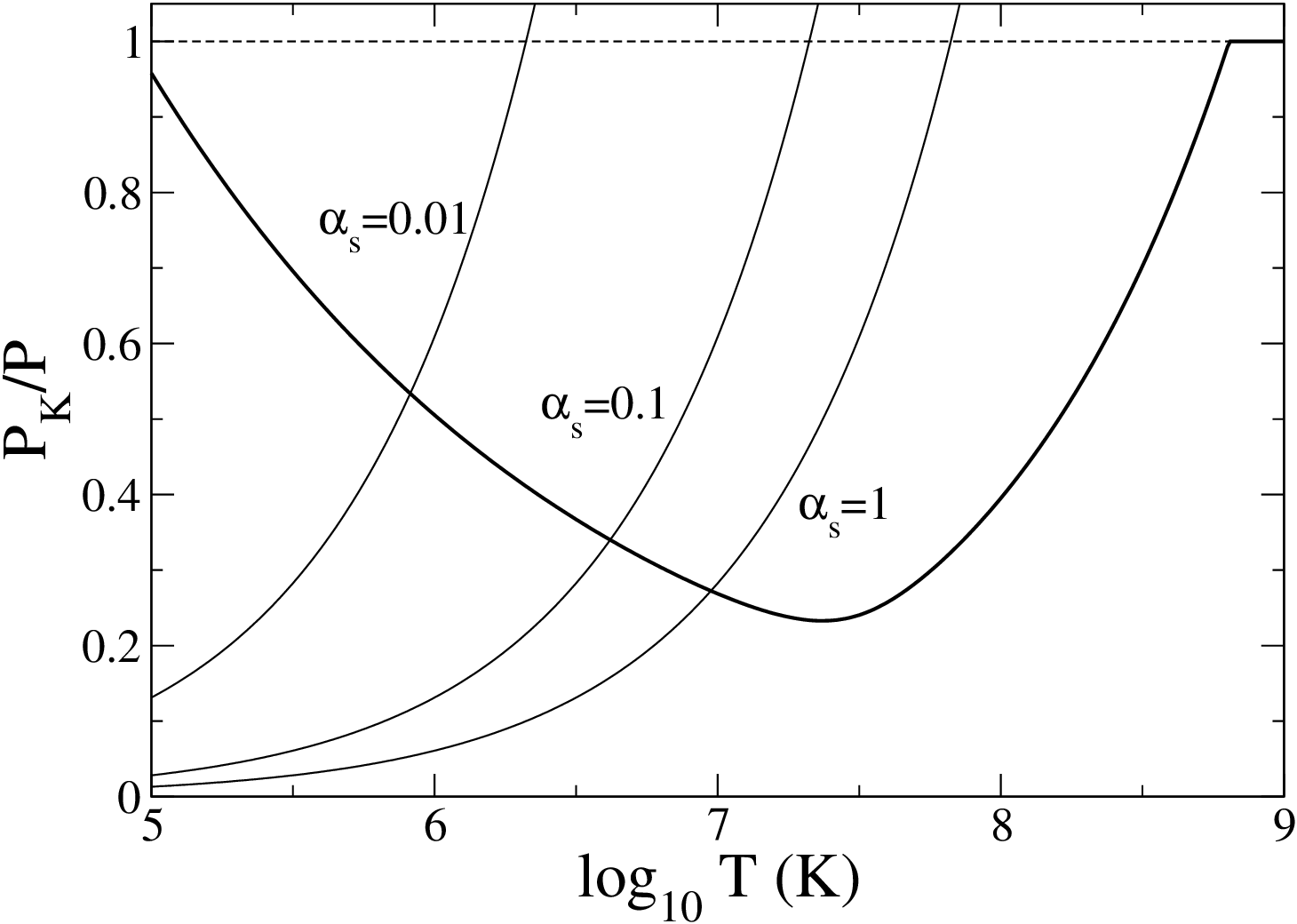} \hfill}
\caption{An illustration of the relevance of the nonlinear contribution 
to the strange star bulk viscosity. We show  the curves in the 
$P-T$ plane corresponding to various values of the saturation amplitude
$\alpha_s$, cf. Eqn~(\ref{nonlin}). Also shown is the main instability
window for a strange star (thick solid line). }
      \label{saturate}
\end{figure}

We can also estimate the amplitude at which the r-mode 
saturates because of the nonlinear contribution to the viscosity. 
To do this (in a suitably simple way that avoids volume 
integration over eigenfunctions etc) we use
\be
{ 3 \over 16 \pi^2} \left(
{m_s^2 \over 3\mu_d T}{\delta r \over R} \right)^2 \approx 
0.36 \alpha^2 P_{-3}^{-4} m_{100}^4 \mu_{400}^2 T_9^2 
\ee
and modify the bulk viscosity 
timescale to 
\be
t_b^{\rm new} = t_b \left[ 1 +  0.36 \alpha^2 P_{-3}^{-4} m_{100}^4 \mu_{400}^2 T_9^2 \right]^{-1}
\label{newbulk}\ee
Assuming that the mode amplitude is large (such that we can neglect the 
first term in the bracket of the denominator), and that 
the shear viscosity is suitably weak, we arrive at a saturation 
amplitude (by setting $\dot{\alpha}=0$ the evolution equation for the mode 
amplitude)
\be
\alpha_s^2 \approx 19.4  \mu_{400}^2 P_{-3}^6 T_8^4 M_{1.4}^{-3} 
\label{nonlin}\ee
Although this is a rough estimate that may be far from the detailed answer it 
shows that the effect could, in principle, be relevant and that nonlinear 
viscosity saturation should be taken seriously for strange stars.

However, as is clear from Figure~\ref{youngevol} a young strange star
never evolves into the regime where the nonlinear contribution to the 
bulk viscosity dominates. Hence, although it is conceptually interesting, 
we do not 
expect this effect to play a role in saturating an
unstable r-mode in a young strange star that enters the instability 
window spinning at or near the breakup limit.


\begin{thebibliography}{10}


\bibitem[\protect\citename{Alcock et al }1986]{alc86}
Alcock C., Farhi E., Olinto A., 1986, {\em Ap. J.} {\bf 310} 261

\bibitem[\protect\citename{Alpar }1987]{alp87}
Alpar, M.A., 1987,   Phys. Rev. Lett. {\bf 58} 2152 

\bibitem[\protect\citename{Andersson \& Kokkotas }2001]{akreview} 
Andersson, N., Kokkotas, K.D., 2001, Int. J. Mod. Phys. {\bf 10} 381 

\bibitem[\protect\citename{Andersson et al }2000]{runaway}
Andersson N., Jones D.I.,  Kokkotas K.D., Stergioulas N., 2000
Ap. J. Lett. {\bf 534} L75

\bibitem[\protect\citename{Andersson \& Comer }2001]{ac01}
Andersson N., Comer G.L., 2001, MNRAS, 328, 1129

\bibitem[\protect\citename{Andersson, Kokkotas \& Schutz }1999]{aks00}
Andersson N., Kokkotas K.D., Schutz B.F., 1999,  Ap. J {\bf 510} 846 

\bibitem[\protect\citename{Andersson, Kokkotas \& Stergioulas }1999]{akst99}
Andersson N., Kokkotas K.D., Stergioulas N., 1999, 
Ap. J {\bf 516} 307 

\bibitem[\protect\citename{Bildsten }1998]{bild}
Bildsten L., 1998 Ap. J. Lett. {\bf 501} 89 

\bibitem[\protect\citename{Bildsten \& Ushomirsky }2000]{bu}
Bildsten L.,  Ushomirsky G., 2000, Ap. J. Lett. {\bf 529} 33

\bibitem[\protect\citename{Colpi \& Miller }1992]{colpi}
Colpi M., Miller J.C., 1992, Ap. J. {\bf 388} 513

\bibitem[\protect\citename{Edwards, Straten \& Bailes }2001]{edwards}
Edwards R.T., van Straten W., Bailes M., 2001, Ap. J. {\bf 560} 365

\bibitem[\protect\citename{Farhi \& Jaffe }1984]{fjaffe}
Farhi E., Jaffe R.L., 1979, Phys Rev D {\bf 30}, 2379

\bibitem[\protect\citename{Glendenning \& Weber }1992]{gw92}
Glendenning N. K., Weber F., 1992, {\em Ap. J.} {\bf 400} 647

\bibitem[\protect\citename{Ho \& Lai }2000]{ho}
Ho W.C.G, Lai D., 2000, Ap. J. {\bf 543} 386

\bibitem[\protect\citename{Jones }2001]{jones}
Jones P.B., 2001, Phys. Rev. Lett. {\bf 86} 1384

\bibitem[\protect\citename{Lattimer et al }1994]{lat}
Lattimer J.M., van Riper K.A., Prakash M., Prakash M., 1994, 
Ap. J. {\bf 425} 802

\bibitem[\protect\citename{Levin }1999]{levin}
 Levin Y., 1999, Ap. J. {\bf 517} 328

\bibitem[\protect\citename{Levin \& Ushomirsky }2000]{lu00}
Levin Y., Ushomirsky, G., 2000, astro-ph/0006028

\bibitem[\protect\citename{Lindblom }2001]{lindrev}
Lindblom L., 2001,  {\em Neutron Star Pulsations and Instabilities} 
pp. 257-276 in
       "Gravitational Waves: A Challenge to Theoretical Astrophysics," Ed: 
V. Ferrari, J.C. Miller, and L. Rezzolla (ICTP, Lecture Notes Series) [preprint astro-ph/0101136]

\bibitem[\protect\citename{Lindblom, Owen \& Morsink }1998]{lom}
Lindblom L., Owen B.J., Morsink S., 1998, Phys. Rev. Lett. 80, 4843 

\bibitem[\protect\citename{Lindblom, Owen \& Ushomirsky }2000]{lou}
Lindblom L., Owen B.J., Ushomirsky G., 2000, Phys Rev D. {\bf 62} 084030

\bibitem[\protect\citename{Lindblom \& Owen }2001]{lo01}
Lindblom L., Owen B.J., 2001, {\em Effect of hyperon bulk viscosity on neutron-star r-modes} preprint astro-ph/0110558

\bibitem[\protect\citename{Lindblom \& Mendell }2000]{lm00}
Lindblom L.,   Mendell G., 2000, Phys. Rev. D {\bf 61} 104003

\bibitem[\protect\citename{Lindblom, Tohline \& Vallisneri }2000]{ltv01}
Lindblom L., Tohline J.E., Vallisneri M., 2001, 
Phys. Rev. Lett. {\bf 86}  1152

\bibitem[\protect\citename{Madsen }1992]{mad92}
Madsen J., 1992,  Phys. Rev. D {\bf 46} 3290

\bibitem[\protect\citename{Madsen }1998a]{mad98}
Madsen J., 1998a,  Phys. Rev. Lett. {\bf 81} 3311

\bibitem[\protect\citename{Madsen }1998b]{madrev}
Madsen J., 1998b, in "Hadrons in Dense Matter and Hadrosynthesis", 
Lecture Notes in Physics, Springer Verlag, Ed. J.Cleymans

\bibitem[\protect\citename{Madsen }2000]{mad00}
Madsen J., 2000,  Phys. Rev. Lett. {\bf 85} 10

\bibitem[\protect\citename{Mendell }2000]{mendell}
Mendell G., 2001, Phys. Rev. D {\bf 64} 044009

 
\bibitem
[\protect\citename{Middleditch et al.\ }2000]
{middle}
Middleditch J., et al.\, 2000, {\em New Astronomy} {\bf 5}  243

\bibitem[\protect\citename{Owen et al }1998]{owen}
Owen B.J., Lindblom L., Cutler C., Schutz B.F., Vecchio A., Andersson N.,
1998, Phys Rev D, {\bf 58}, 084020.

\bibitem[\protect\citename{Page et al }2000]{page}
 Page D., Prakash M., Lattimer J.M.,  Steiner A.W., 2000,
 Phys. Rev. Lett. {\bf 85} 2048 

\bibitem[\protect\citename{Papaloizou \& Pringle }1978]{ppinst}
 Papaloizou J.,  Pringle J.E., 1978, MNRAS {\bf 184} 501

\bibitem[\protect\citename{Pizzochero }1991]{piz91}
Pizzochero P. M., 1991,  Phys. Rev. Lett. {\bf 66} 2425

\bibitem[\protect\citename{Rezzolla, Lamb \& Shapiro }2000]{rezz}
 Rezzolla L., Lamb F.K., Shapiro S.L., 2000, Ap. J. Lett. {\bf 531} 139

\bibitem[\protect\citename{Spruit }1999]{spruit}     
Spruit H.C., 1999, Astron. Astrophys. {\bf  41} L1

\bibitem[\protect\citename{Stergioulas \& Font }2000] {sf00}
Stergioulas N., Font J.A., 2001, Phys. Rev. Lett. {\bf 86}  1148

\bibitem[\protect\citename{Usov }2001]{usov01}
Usov, V.V., 2001, Phys. Rev. Lett. 87, 1101

\bibitem[\protect\citename{Wagoner }1984]{wag}
Wagoner R.V., 1984, Ap. J. {\bf 278} 345

\bibitem[\protect\citename{Watts \& Andersson }2001]{watts}
Watts A.L., Andersson N., 2001 {\em The spin evolution of nascent neutron stars} preprint astro-ph/0110573

\bibitem[\protect\citename{Witten }1984]{witten}
Witten E., 1984, Phys. Rev. D {\bf 30} 272

\bibitem[\protect\citename{Yuan \& Zhang }1999]{yz99}
Yuan Y. F., J. L. Zhang, 1999,  Astron. Astrophys. {\bf 344} 371 

\end{thebibliography}
\end{document}